\documentclass[%
 amsmath,amssymb,
preprint,%
superscriptaddress,%
]{revtex4-1}

\usepackage{graphicx}
\usepackage{subfigure}
\usepackage{dcolumn}
\usepackage[normalem]{ulem}
\usepackage{bm}
\usepackage{verbatim}
\usepackage{xcolor}
\usepackage{float}

\begin{document}

\title
{Kinetic Monte Carlo Simulations of the Effect of the Exchange Control Layer Thickness in CoPtCrB/CoPtCrSiO Granular Media}
\author{Ahmad M. Almudallal}
\affiliation{Department of Physics and Physical Oceanography,
Memorial University, St. John's, NL, Canada A1B 3X7} 

\author{J. I. Mercer}
\affiliation{Department of Computer Science, 
Memorial University of Newfoundland, St. John's, NL, Canada A1B 3X7 }

\author{J. P. Whitehead}
\affiliation{Department of Physics and Physical Oceanography,
Memorial University, St. John's, NL, Canada A1B 3X7} 

\author{M. L. Plumer}
\affiliation{Department of Physics and Physical Oceanography,
Memorial University, St. John's, NL, Canada A1B 3X7} 

\author{ J. van Ek}
\affiliation{7206 Meadowdale Dr., Niwot, CO, 80503, USA}
\date{\today}

\begin{abstract}
A hybrid kMC/LLG algorithm is used to simulate experimental MH hysteresis loops for dual layer ECC media. The calculation of the rate coefficients and difficulties arising from low energy barriers, a fundamental problem of the kMC method, are discussed and the methodology used to treat them in the present work is described. The results from simulations are compared with experimental data on dual layer ECC CoPtCrB/CoPtCrSiO media. A quantitative relationship between the thickness of the exchange control layer and the effective exchange constant between the layers is demonstrated. 

\end{abstract}

\maketitle


The modeling of MH hysteresis loops of highly anisotropic materials at experimentally relevant temperatures and sweep rates based on standard micromagnetic models represents a significant challenge. Such materials typically show a strong sweep rate dependence  as the grain reversal process is dominated by thermal activation involving large energy barriers ($\Delta E \gg k_BT$). As a consequence, the usual stochastic Landau-Lifshitz-Gilbert (sLLG) method~\cite{plumer,brown}, that is widely used in micromagnetics to simulate magnetic materials at finite temperature, is simply not feasible because of the long time scales involved. Perpendicular recording media (PRM) used in hard disc drives (HDD) represent an important class of such materials. By characterizing the relaxation process as a sequence of quasi-equilibrium states separated by thermally assisted grain reversals, kinetic Monte Carlo (kMC) provides an alternative approach to sLLG that can be applied to study such materials~\cite{chantrell, kanai, lu, charap, hovorka, parker, chaves, tan, chureemart}.

This work presents results based on a hybrid kMC/LLG formalism, that has been applied to single layer PRM~\cite{falPRB}, to simulate MH loops of dual layer exchange coupled composite (ECC) media~\cite{ahmadPRB}. As shown in earlier work the parameters obtained from fitting experimental MH loops for ECC media from simulations using the kMC/LLG formalism at experimental sweep rates differ significantly from those obtained using stochastic LLG that are limited to sweep rates several orders of magnitude greater than the experimental sweep rates~\cite{falAPL}. The simulation results presented here are compared with published experimental VSM studies that examine the effect of the thickness of the exchange control layer (ECL)  in CoPtCrB/CoPtCrSiO ECC media~\cite{MHvsm}. Such measurements and the associated simulation studies represent an important tool in determining the material parameters  in order to optimize the competing demands of the areal bit density, writability and stability of ECC media.  

The simulations are based on a model in which each of the grains is represented by two exchange-coupled Stoner-Wohlfarth particles, which we label as $a$ (cap layer) and $b$ (oxide layer). The volumes of the $a$ and $b$ layers comprising grains, and the cross sectional area of the interface separating them, are denoted by $v_a$, $v_b$ and $A$, respectively. The energy of the $k^\mathrm{th}$ grain may then be written as
\begin{align}
E_k =& - K^k_a v_a\left(\hat m_a^k \cdot \hat n_a^k\right)^2 -  K^k_b v_b\left(\hat m_b^k \cdot \hat n_b^k\right)^2 -IA\; (\hat m_a^k \cdot \hat m_b^k)\nonumber\\ 
&- \vec M^k_a \cdot\bar \Gamma^{k}_{aa} \cdot \vec M^k_a  - \vec M^k_b \cdot \bar\Gamma^{k}_{bb} \cdot \vec M^k_b -\vec M^k_a \cdot \bar{\Gamma}^{k}_{ab} \cdot \vec M^k_b\nonumber \\
&-\sum_{k\ne k'} \left(\vec H^a_{kk'} \cdot \vec m^k_a + \vec H^b_{kk'} \cdot \vec m^k_b\right)  - \vec H\cdot \left(\vec m^k_a+\vec m^k_b \right) 
\end{align}
where $\vec H=H\hat z$ is the perpendicular applied magnetic field, $K^k_l$, $\hat n^k_l$ and $\vec M^k_l$ are the anisotropy constant, anisotropy axis and the magnetization vector in each of the layers ($l\in\{a,b\}$), respectively, with $\vec m^k_l = \vec M^k_l v_l$ and $\hat m^k_l = \vec M^k_l/ M^k_l$.  $\bar{\Gamma}^{k}_{ab}$ denotes the magnetostatic tensor between the layers $a$ and $b$, with $\bar{\Gamma}^{k}_{aa}$ and $\bar{\Gamma}^{k}_{bb}$ denoting the magnetostatic shape anisotropy for the layers $a$ and $b$. $\vec H^l_{kk'}$ is the interaction field (exchange plus magnetostatic) acting on the $l^\mathrm{th}$ layer ($l \in \{a,b\}$) of the $k^\mathrm{th}$ grain due to the $k^{\prime\,\mathrm{th}}$ grain, and $IA$ represents the exchange interactions between the $a$ and $b$ layers within a single grain, where $A$ denotes the cross sectional area of the ECL and we refer to $I$ as the interlayer coupling constant.

The simulation of the MH loops using the hybrid kMC/LLG approach is described in some detail for both single and dual layer materials in Refs.~\onlinecite{falPRB,ahmadPRB}. The simulation begins with the system in a fully saturated state with an applied field $H_0$. The saturated state is then input into an LLG ($T=0$) simulation and relaxed to a local minimum energy state, which we denote by $\mathcal{S}_0$, and the interaction field $\sum_{k'\ne k} \vec H^l_{k'k}$ is calculated for each layer ($l \in \{a,b\}$) in each grain ($1\le k \le L\times L$). The energy of each grain may then be calculated as a function of $\hat m^k_l$ and the minimum energy states for each grain is determined. For grains with more than a single minimum energy state we compute the rate constants $r^k_{i\to j}$ between each pair of minimum energy states based on the Arrhenius-N\'{e}el expression  
\begin{align}
r^k_{i\to j} = f^k_{ij} \exp\left(-\frac{\Delta E^k_{ij}}{k_BT}\right) 
\end{align}
where the ordered pairs $\{ij\}$ label the minimum energy states in the $k^\mathrm{th}$ grain connected by a minimum energy path (MEP).  $\Delta E^k_{ij}$ and $f^k_{ij}$ denote the energy barrier and attempt frequency separating the initial minimum energy states $i$ from the final state $j$. From the rate constant, the set of wait times $t^k_{i\to j} \log x$, where $x$ is a uniformly distributed random number between 0 and 1, is calculated for each of the grains and the grain with the minimum wait time $t_R = \min[t^{k}_{i\to f}]$ determined. Denoting the index of this grain as $k_R$, the quantity $t_R$ is referred to as the time to first reversal and essentially determines the time at which the grain $k_R$ undergoes a reversal from some initial state $i$ to some final state $f$. If $t_R$ is less than some user specified time interval $\Delta t$ then a new state $\mathcal{S}_1$ at time $t_1 = t_R$ is constructed in which the grain at $k=k_R$, initially in state $i$, is replaced with the grain in state $f$ and the system is allowed to relax to some new local equilibrium with $H= H_1 =H_0 - R\,t_R$, where $R=|dH/dt|$ denotes the sweep rate. If, on the other hand, $t_R \ge \Delta t$, then the $k^\mathrm{th}$ grain remains in the state $i$ and $\mathcal{S}_1 = \mathcal{S}_0$ and the system is allowed to relax to some new local equilibrium with $H=H_1= H_0 - R\,\Delta t$. This process is repeated generating a sequence of states $\{\mathcal{S}_0, \mathcal{S}_1\dots \}$ at times $\{t_0,t_1 \dots\}$ until the normalized magnetization $M(S_n) < -M_s$, where $M_s < 1$ is some nominal value used to define saturation. 
\begin{table}[t!]
\begin{center}
\begin{tabular}{| c | c | c | c |}
\hline
\bf{Cap Layer} & $M_a$ &  $K_a$ &  $A_a$  \\\hline
Experimental & $425$  & $2.20 \times 10^6$ &  ---  \\\hline
Simulation      & $450$ & $2.10 \times 10^6 $ & $2.0 \times 10^{-6}$ \\\hline\hline
\bf{Oxide layer} &  $M_b$ &  $K_b$ &  $A_b$ \\\hline
Experimental &  $385$ &  $3.10 \times 10^6$  & ---  \\\hline
Simulation      & $385$ & $3.05 \times 10^6 $  & $0.018 \times 10^{-6}$ \\\hline\hline 
 \bf{Units} &  (emu/cc) & (erg/cc) & (erg/cm)\\\hline
\end{tabular}
\caption{Parameters used in modelling the cap layer ($M_a$, $K_a$ and $A_a$) and the oxide layer ($M_b$, $K_b$ and $A_b$), where $M$, $K$, and $A$ represent the magnetiztion, anisotropy and exchange stiffness.  Experimental values were extracted from Ref.~\cite{MHvsm}.}\label{tab:Parameters}
\end{center}
\end{table}
The application of kMC to the case of single layer PRM is relatively straightforward~\cite{falPRB}. Each grain typically has a maximum of two energy minima and the location of the minima, the saddle points connecting them and the calculation of the associated rate constants can be performed analytically. The dual layer case on the other hand presents a number of challenges. The most obvious is the fact that it requires the numerical determination of the energy minima, the saddle points connecting them and the associated rate constants  for each grain at each kMC step. This is a formidable task and the efficiency and stability of the algorithm used determines the feasibility of the kMC method. 

In the present work we discretize the unit sphere that describes the state-space of a single spin by triangulating the surfaces of an octahedron inscribed by the unit sphere and projecting the vertices onto the surface of the sphere. The resultant polyhedron is a 3-polytope consisting of approximately 144 vertices enumerated by the index $\sigma$. 
The discretized states of a dual layer grain are then given by the cartesian product $(\sigma_a, \sigma_b)$. The location of minimum energy states of the ECC grains are obtained from a steepest descent algorithm and is then applied to each of the points $x =(x_a(\sigma),x_b(\sigma))$. Points that coincide with a certain tolerance are then merged to give the minimum energy states of the grain. All the paths between any two minimum energy states can then be constructed from the edges of the polyhedron and the minimum energy path on the lattice determined using a modified Bellman-Ford algorithm~\cite{bellmanFord}. This provides an initial estimate of the location of the saddle points that can be further refined by fitting the energy surface to a quartic at $x =(x_a(\sigma),x_b(\sigma))$. From this the energy barrier separating the two states and the attempt frequency may be calculated~\cite{ahmadPRB}.

In addition to the complexity of the energy landscape of dual layer ECC grains, and the computational challenges they present in determining the rate coefficients, the variation in the energy barriers can give rise to a range of rate constants that span several orders of magnitude. The presence of these large rate constants associated with transitions between states separated by low energy barriers does not contribute significantly to the hysteresis loops observed at experimental measurement times as the rapid fluctuations between these transient states mean that they quickly equilibrate. However, while it is the low frequency transition between states separated by large energy barriers that essentially controls the relaxation processes at experimental sweep rates the presence of these high frequency transitions nevertheless can severely limit the time scales that can be simulated using the kMC algorithm. This is an outstanding problem with the kMC algorithm~\cite{kmcReview} and has proven to be a significant factor in its application to dual layer ECC media. 

To deal with this problem, the kMC/LLG code used here is programmed to detect the onset of high frequency fluctuations by identifying grains that return to their initial state after $N$ kMC steps (in the present code $N$ is set to 32). When these high frequency fluctuations are detected, minimum energy states that are connected by an energy barrier less than some predetermined threshold are combined into clusters. The kMC is then reformulated in terms of a combination of minimum energy states and clusters~\cite{ahmadPRB}. If the fluctuations persist, then the threshold is increased by a factor of 2 until such time as the high frequency fluctuations are suppressed and the time between successive kMC steps is appropriate. The threshold is then gradually reduced until the high frequency fluctuations reappear. This adaptive scheme for suppressing the high frequency fluctuations that do not contribute to the long time system dynamics allows us to model complex systems governed by disparate energy barriers both efficiently and accurately~\cite{ahmadPRB}. 

\begin{figure}[ht]
\centering
 \includegraphics[width=0.45\textwidth]{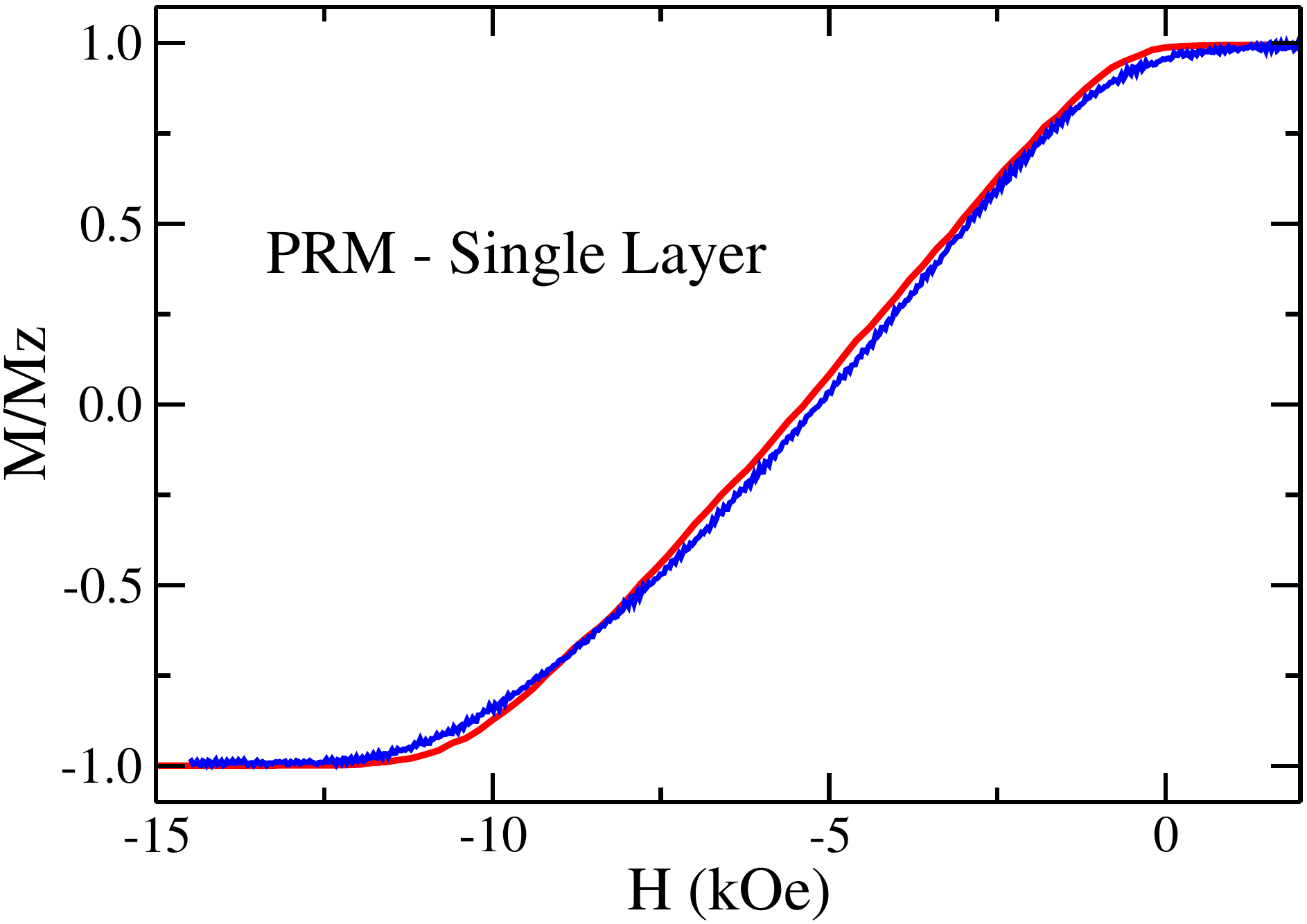}
\caption{\label{fig:singleLayerMH} Normalized magnetization as a function of applied field for the $14 \,\mathrm{nm}$ oxide layer, CoPtCrSiO, at a sweep rate $R = 7.5 \,\mathrm{kOe/s}$ from experimental data (blue) and kMC simulations (red).}
\end{figure}

In the present model, grains are arranged on a $L\times L$ square lattice with periodic boundary conditions (where $L=32$). The anisotropy constant, anisotropy axis and the magnetization are assigned grain-to-grain Gaussian distributions of values. Oxide and cap layers were $14 \,\mathrm{nm}$ and $4 \,\mathrm{nm}$ thick, respectively, as used in the experimental study~\cite{MHvsm}. Lateral grain dimensions were $7 \,\mathrm{nm}$ $\times$ $7 \,\mathrm{nm}$. There was no spacer layer in the model.  A sweep rate of $R=7.5\,\mathrm{kOe/s}$ at temperature $T=300 \text\,{K}$ was used in both the simulation and experimental studies. The material parameters used in the simulations that reproduced well the experimental data are shown in Table~\ref{tab:Parameters} and were determined as follows. Assumed values for the exchange stiffness parameter, denoted as $A_a$ and $A_b$, were guided by previous modeling results on generic ECC media for the cap and oxide layers, with moderate and weak coupling, respectively~\cite{falAPL}. Experimental values for $M_a$, $M_b$, $K_a$ and $K_b$ (see Table~\ref{tab:Parameters}) served as a starting point for these parameters in the MH loop fitting procedure.  To simplify fitting the MH loops for the dual layer media, we first fit the kMC loop to the experimental results for only the oxide layer. Best results were achieved using the magnetization and anisotropy values indicated in Table~\ref{tab:Parameters}, along with a 10\% variance in $K_b$ and $M_b$ as well as a $4^\circ$ variance in anisotropy axis direction about $z$.  The outcome of this procedure is shown in Fig.~\ref{fig:singleLayerMH} and illustrates excellent agreement between simulation and experimental results.

\begin{figure*}[ht]
\subfigure{\centering\includegraphics[height=4.0cm]{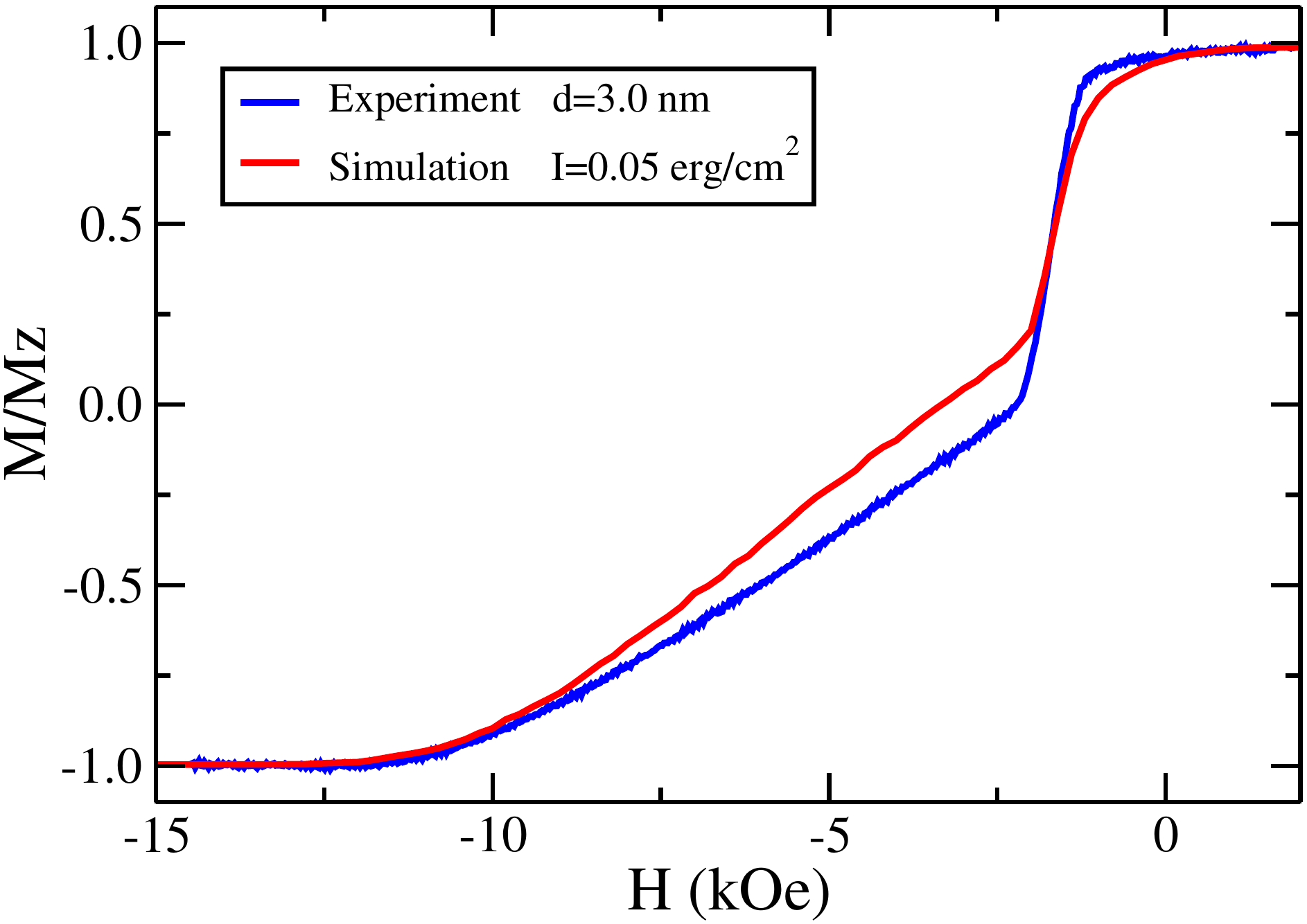}}
\subfigure{\centering\includegraphics[height=4.0cm]{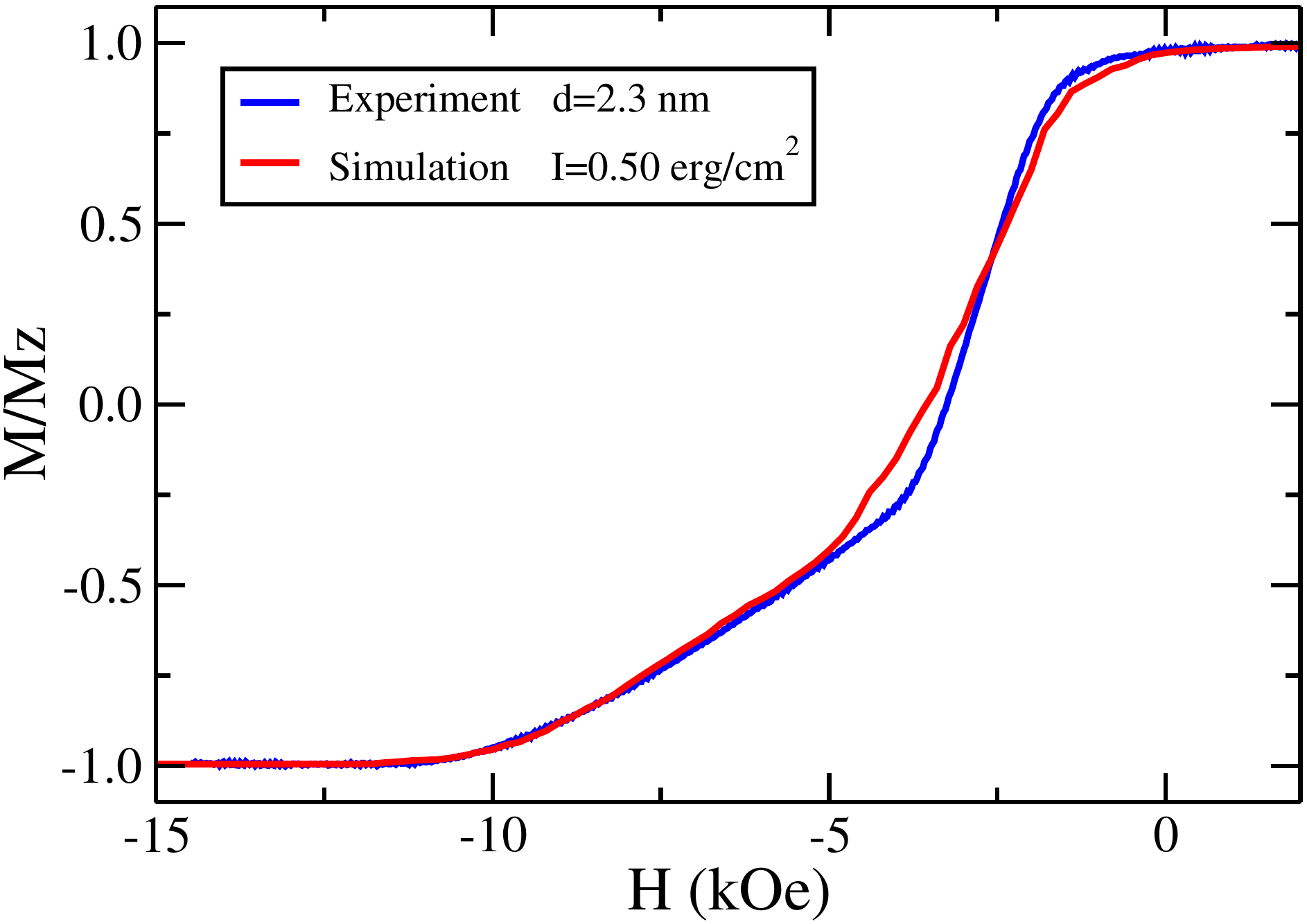}}
\subfigure{\centering\includegraphics[height=4.0cm]{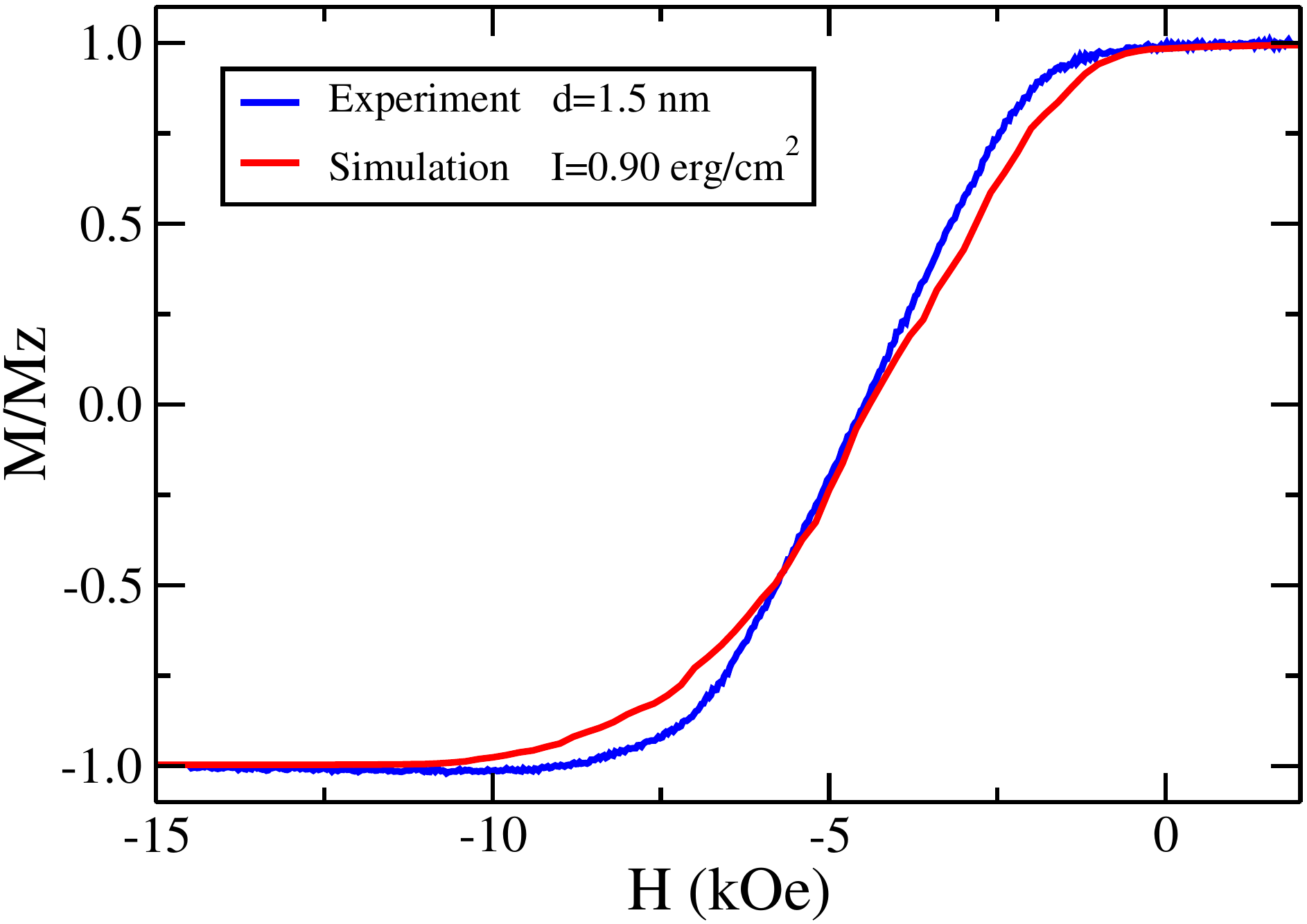}}
\subfigure{\centering\includegraphics[height=4.0cm]{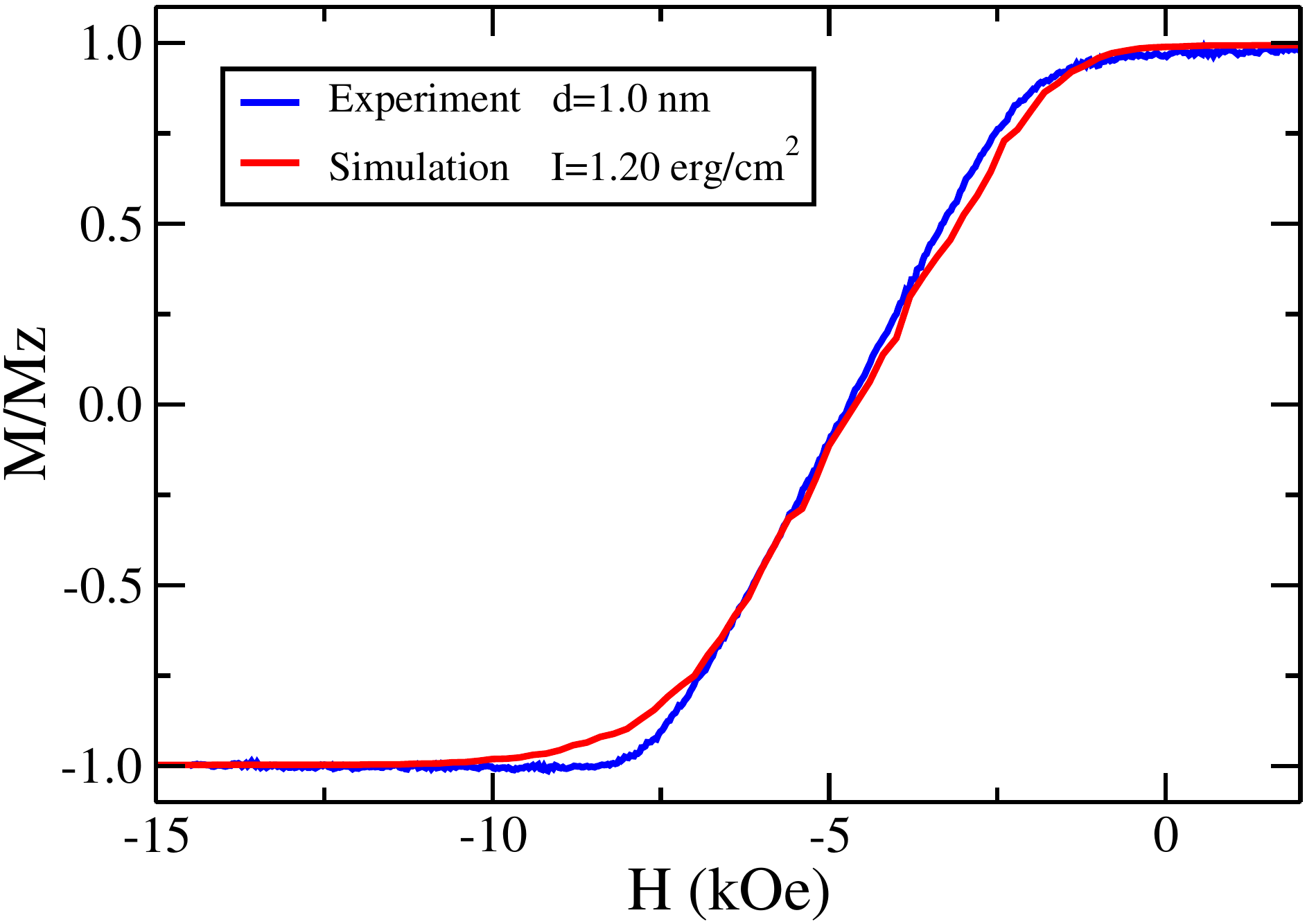}}
\subfigure{\centering\includegraphics[height=4.0cm]{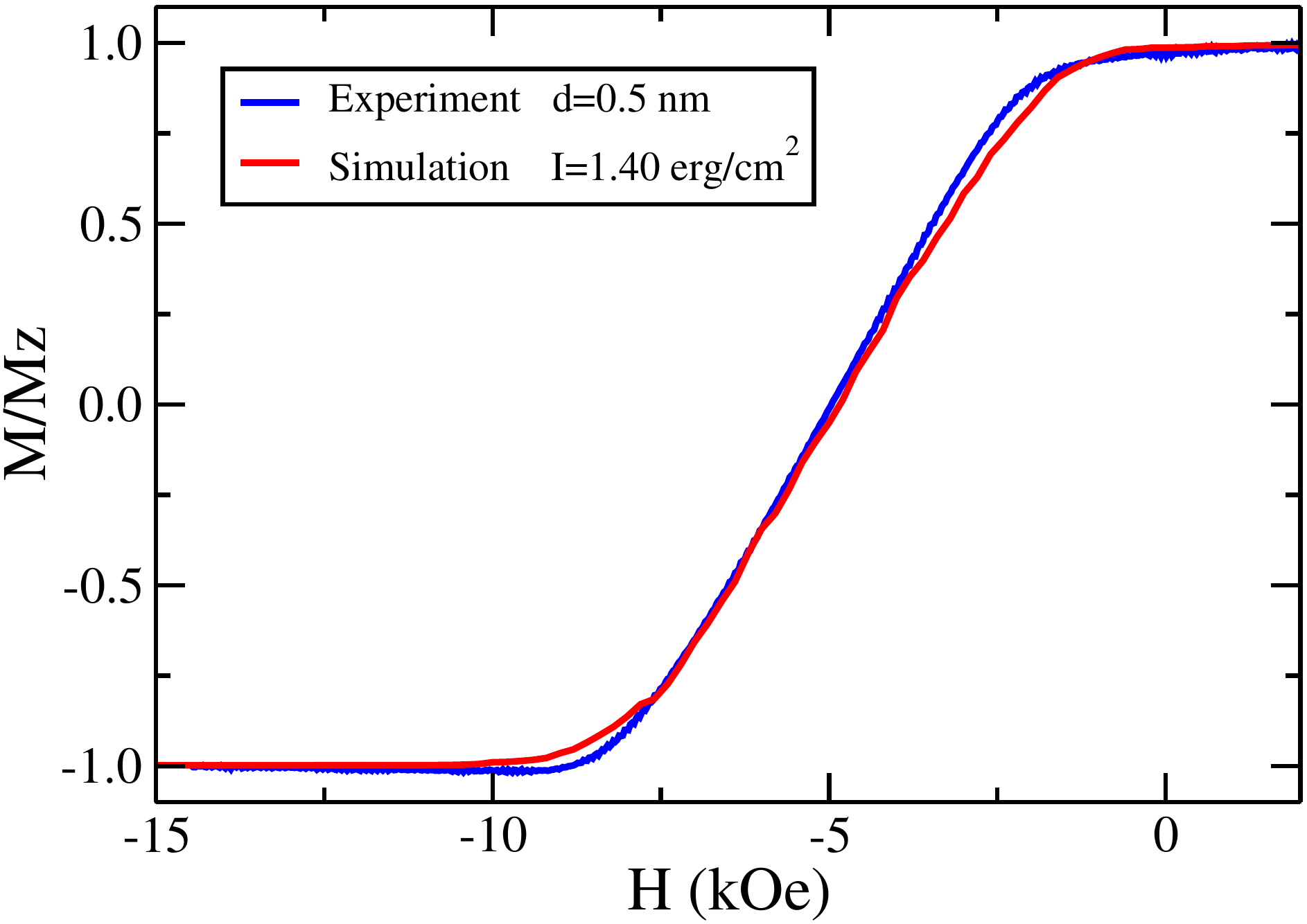}}
\subfigure{\centering\includegraphics[height=4.0cm]{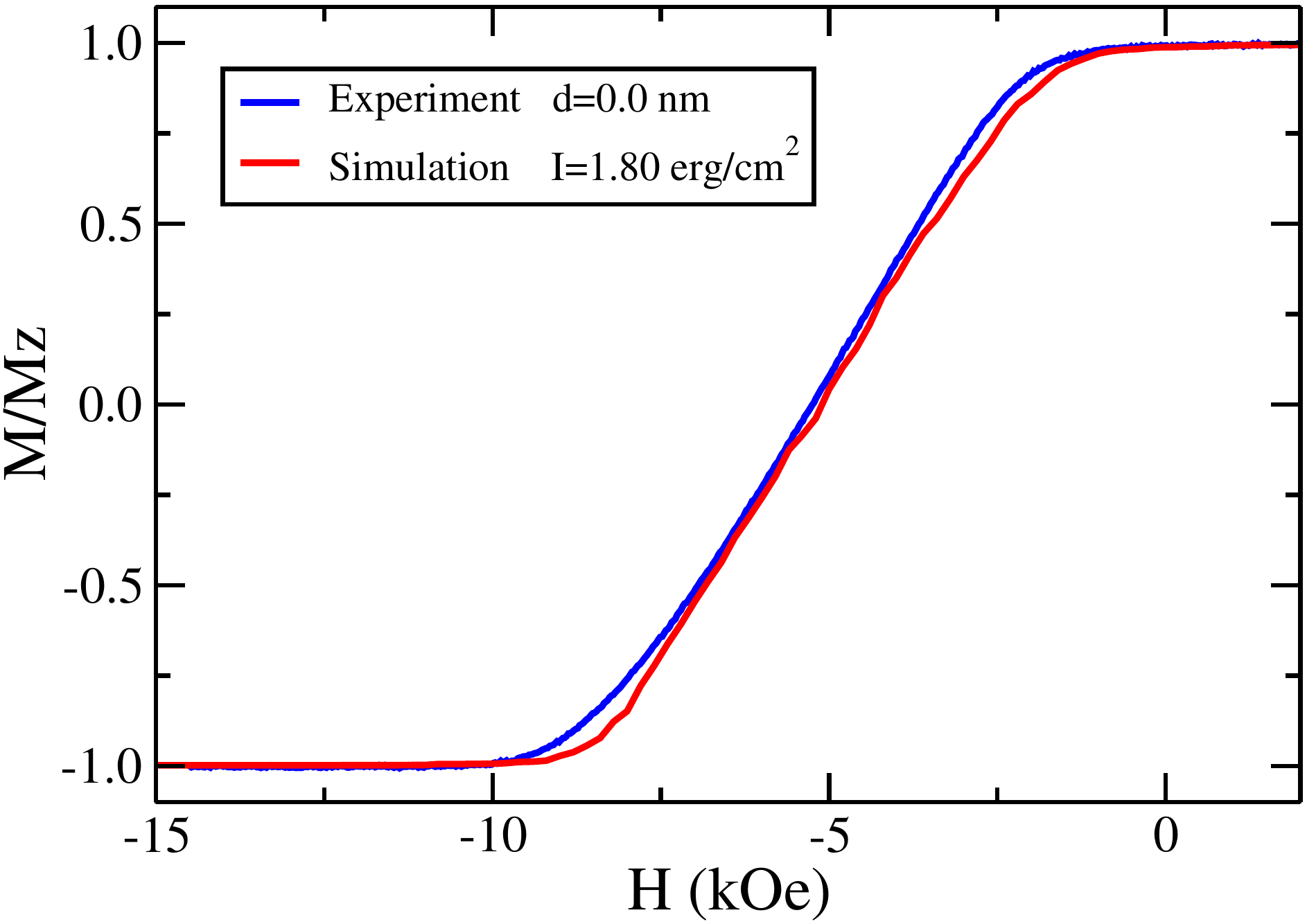}}
\caption{\label{fig:dualLayerMH} MH loops for dual layer  ECC CoPtCrB/CoPtCrSiO media for a sweep rate of $R = 7.5 \,\mathrm{kOe/s}$ calculated from experimental data (blue) for several values of $d$ together with the corresponding results from kMC simulations (red) for several values of $I$.}
\end{figure*}

\begin{figure}[ht]
\centering
 \includegraphics[width=0.45\textwidth]{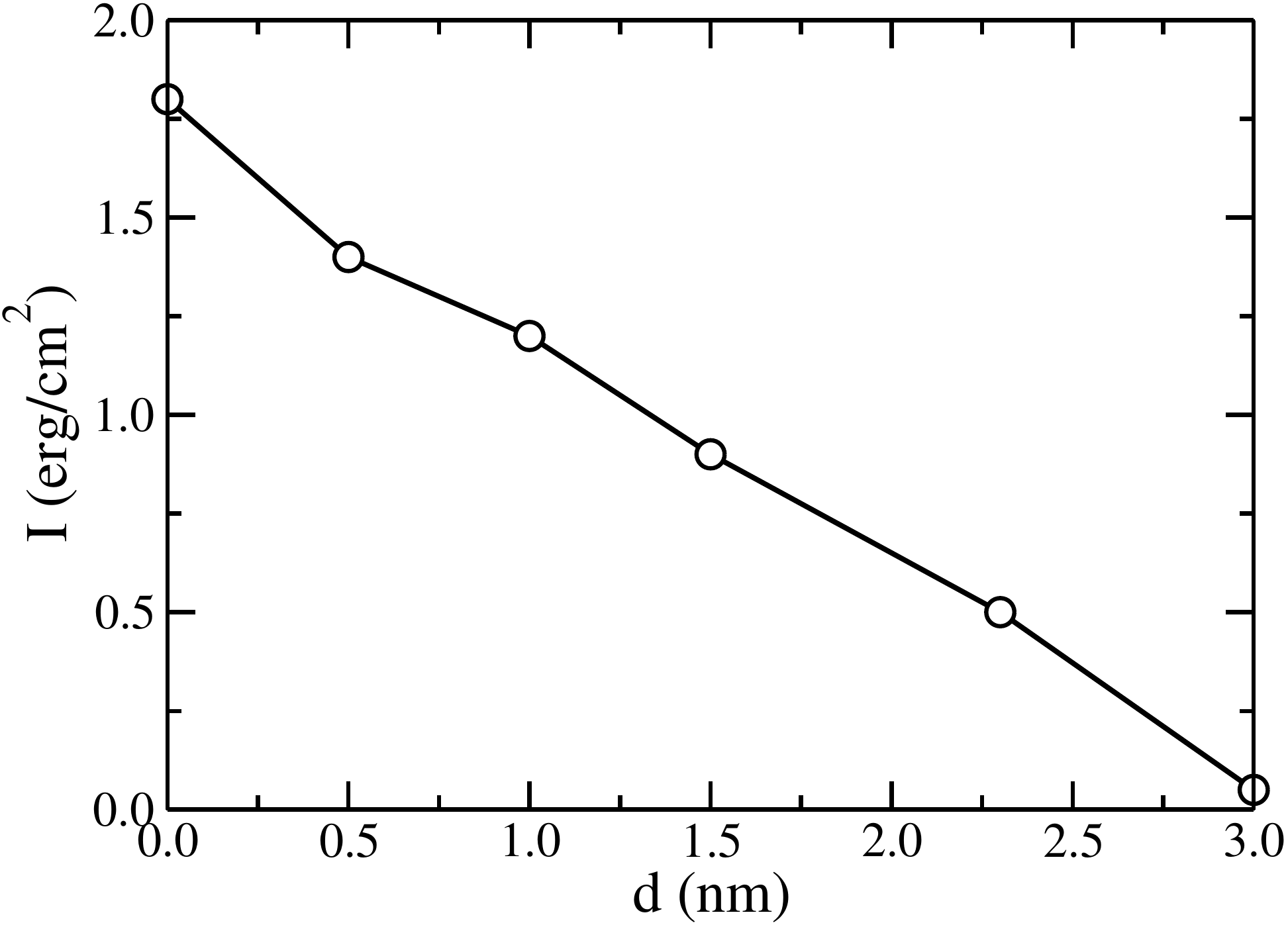}
\caption{\label{fig:Ivsd} Relationship between the interlayer coupling constant $I$ and the ECL thickness $d$ obtained from fitting the simulation results with experiment shown in Fig.~\ref{fig:dualLayerMH}.}
\end{figure}

Having established model parameters for the oxide layer, the hybrid kMC/LLG algorithm was then applied to best reproduce the experimental MH loops in the dual layer case~\cite{MHvsm}. Here, the cap layer magnetization and anisotropy values were refined by best fitting simulation results to the weakly coupled experimental data with a spacer thickness of $d=3.0 \,\mathrm{nm}$.  For this layer, a larger 20\% variance in $K_a$ and $M_a$ (along with the same $4^\circ$ variance in anisotropy axis) were found to best reproduce the data. For all of the other cases of different spacer thickness $d$, only the value of the exchange parameter $I$ was adjusted to achieve the best overall fit between the simulations and the experimental data. The parameter fitting was done by eye based on a single kMC run for each value of $d$. Once a reasonable fit had been obtained a series of 10 kMC runs were performed for each parameter set and averaged over to produce the results presented in Fig.~\ref{fig:dualLayerMH}. In Fig.~\ref{fig:Ivsd}, the resulting values of the exchange parameter $I$ plotted against the ECL thickness $d$ and shows a roughly linear decrease of $I$ as a function of increasing $d$. To our knowledge this is the first calculation that establishes a quantitative relationship between the experimental ECL thickness and the strength of the interlayer exchange in ECC media.

\begin{figure}[htp]
\centering
 \includegraphics[width=0.45\textwidth]{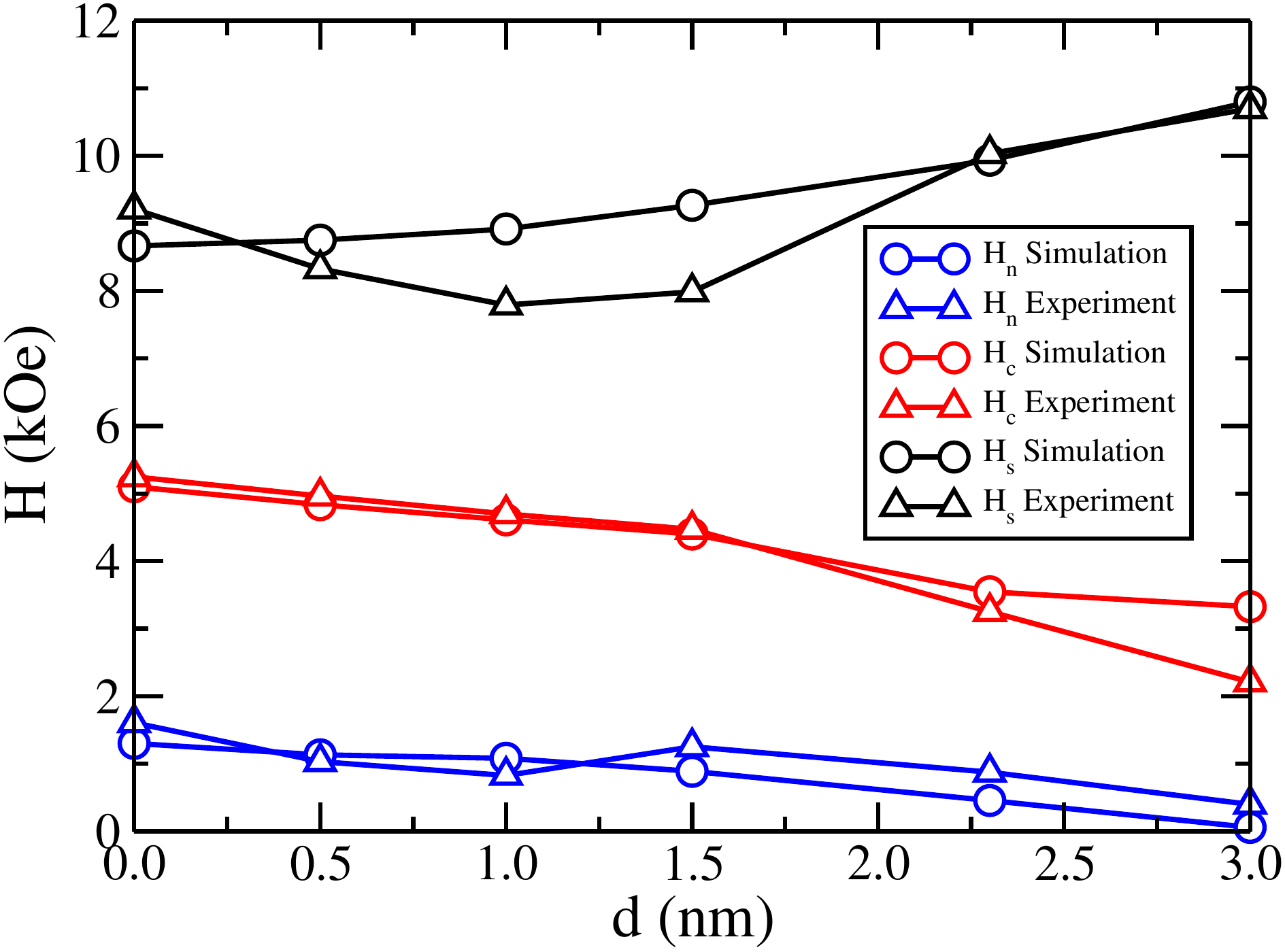}
\caption{\label{fig:Hncs_d}  Nucleation, coercive, and saturation fields $H_n$, $H_c$ and $H_s$ as a function of $d$ for both experimental and kMC simulation. The value of $d$ for the kMC simulations is inferred from Fig.~\ref{fig:dualLayerMH}}
\end{figure}

The MH loops obtained from the simulations shown in Fig.~\ref{fig:dualLayerMH} give good quantitative agreement with the experimental data, and successfully capture their essential features. Figure~\ref{fig:Hncs_d} presents a comparison of the nucleation, coercive, and saturation fields $H_n$, $H_c$, and $H_s$, respectively, estimated from experimental and  simulation results, with the definitions $M(H_n) = 0.95$, $M(H_c) = 0.0$ and $M(H_s) = -0.95$ respectively.  The plots show that while in the case of the nucleation and coercive fields the agreement between the simulation and experiment is quite good, the comparison is less satisfactory in the case of the saturation field. In particular, the experimental values exhibit a well defined minima in $H_s$ for $d \approx 1.0\,\mathrm{nm}$, a feature that is absent in the simulation results. Examining the MH loops presented in Fig.~\ref{fig:dualLayerMH} the discrepancy between the experimental values of $H_s$ and those determined from the simulations is due in part to the shape of the tail of the MH curve in the region $H \approx H_s$. Fig.~\ref{fig:dualLayerMH} also shows that the simulation results underestimate the drop in the magnetization observed at $H \approx 2\,\mathrm{kOe}$ for $d = 0.05\,\mathrm{nm}$ case. 


\begin{figure}[ht]
\centering
\includegraphics[width=0.45\textwidth]{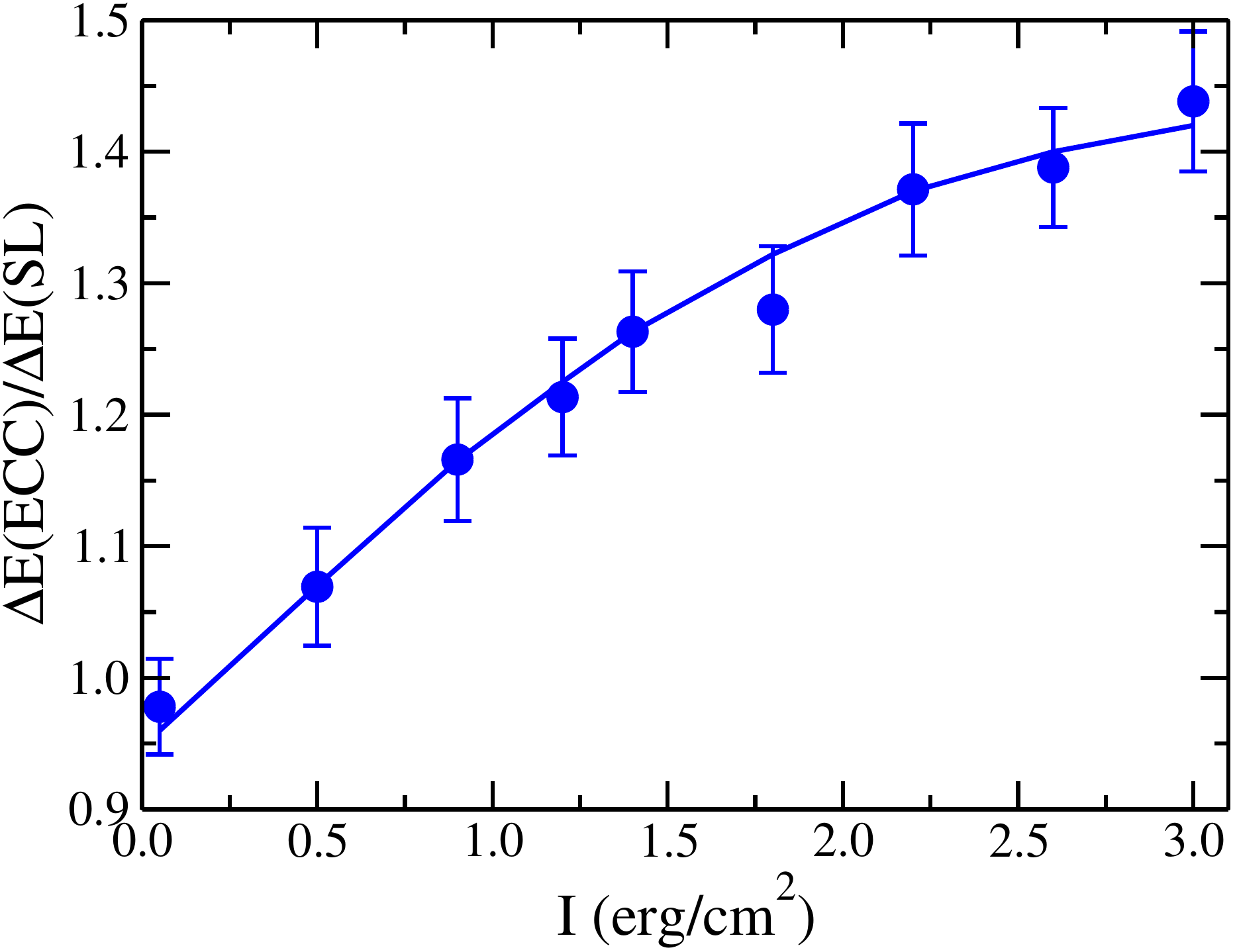}
\caption{\label{fig:DeltaE_I}  Average energy barrier $\langle \Delta E^k_{if} \rangle$ calculated for the dual layer system $\Delta E(\mathrm{ECC})$ normalized by the corresponding single layer $\Delta E(\mathrm{SL})$ result as described in the text for a system allowed to relax to equilibrium from the fully saturated state for $H=0$.}
\end{figure}

\begin{figure}[ht]
\centering
\includegraphics[width=0.45\textwidth]{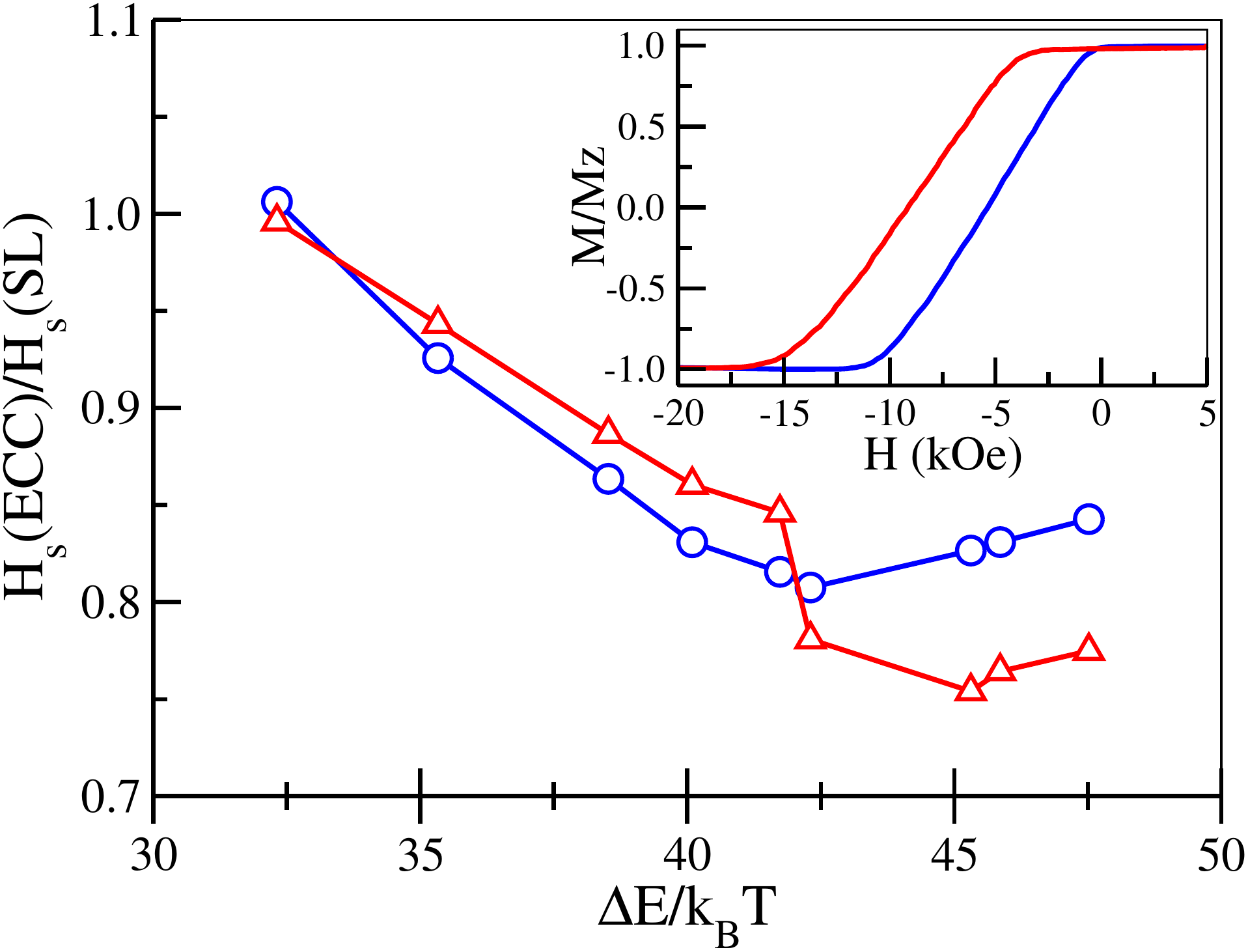}
\caption{\label{fig:Hs_DeltaE} Normalized saturation field $H_s(\mathrm{ECC})/H_s(\mathrm{SL})$ obtained from simulations plotted as a function of $\Delta E/k_BT$  for $R=7.5\,\mathrm{kOe/s}$  (blue curve) and $R=10^7\,\mathrm{kOe/s}$ (red curve). Inset shows a plot of the MH loops obtained from simulations on a single layer for $R = 7.5\,\mathrm{kOe/s}$ (blue curve) and $R = 10^7\,\mathrm{kOe/s}$ (red curve).}
\end{figure}

The distribution of energy barriers for the case of a system allowed to relax to equilibrium from the fully saturated state for $H=0$ has also been calculated for both the dual and single layer media. This is of particular interest as the distribution of energy barriers at zero field is often taken as a measure of the stability of the media with respect to thermal fluctuations. We have calculated the average value $\Delta E=\langle \Delta E^k_{if} \rangle$ over an extended range of the interlayer coupling constant $0 \le I \le 3.0\,\mathrm{erg/cm^2} $, where $\Delta E^k_{if}$ denotes the minimum energy barrier separating the initial state $i$  from all possible final states $f$ of the $k^\mathrm{k}$ grain, with the average $\langle\dots\rangle$ denoting the average over all grains.  The $\Delta E$ values, normalized with respect to the corresponding average calculated for the single layer case, are plotted in Fig.~\ref{fig:DeltaE_I}. The data show the average relative energy barrier increasing with increasing interlayer coupling constant $I$ thereby increasing stability of the grains to thermally activated reversal. We note that  $\Delta E$ does not appear to have reached its saturation value corresponding to a completely coherent rotation of the grains~\cite{ahmadPRB}. Combining the results presented in Figs.~\ref{fig:Hncs_d} and \ref{fig:DeltaE_I} we see that the simulation indicates that, up to $I=1.8\,\mathrm{erg/cm^2}$ ($d=0\,\mathrm{nm}$), $H_s$ decreases with increasing $I$ (decreasing $d$) while the average energy barrier $\Delta E$ increases. This is shown explicitly in Fig.~\ref{fig:Hs_DeltaE} in which we plot $H_s$ vs $\Delta E/k_BT$ ($T=300\,\mathrm{K}$). The first six points ($ 32.5 \le \Delta E/k_BT \le 42.5$), corresponding the range of $I$ used to fit the experimental data shown in Fig.~\ref{fig:dualLayerMH} and show $H_s$ decreasing with increasing $\Delta E/k_BT$. Interestingly, however, for higher values of $I$ the simulations (not shown here) also show  $H_s$ increasing with increasing $I$, indicating that the minimum in $H_s$ observed in the experimental data at $d\approx 1.0\,\mathrm{nm}$ ($I \approx 1.2\,\mathrm{erg/cm^2}$)  in Fig.~\ref{fig:Hncs_d} also appears in the simulations but at a somewhat higher value of $I\approx 1.8 \,\mathrm{erg/cm^2}$. 

While $MH$ loops measured using VSM are a primary means of characterizing the properties of magnetic materials, the typical sweep rates are several orders of magnitude less than the effective sweep rates that pertain to the write process in HDD media. We have extended our studies by analyzing MH loops for the above parameter set at $R=10^7 \,\mathrm{kOe/s}$ using sLLG~\cite{maglua} for both the single and dual layer media. MH loops calculated for single layer of CoPtCrSiO for a sweep rate $R = 7.5 \,\mathrm{kOe/s}$ and $10^7 \,\mathrm{kOe/s}$ are shown in Fig.~\ref{fig:Hs_DeltaE} and show a significant increase in $H_s$ for the higher sweep rate.  Results for normalized values of $H_s(\mathrm{ECC})/H_s(\mathrm{SL})$ are plotted in Fig.~\ref{fig:Hs_DeltaE} as a function of ${\Delta E}/k_BT$. While the value of $H_s$ calculated for both dual and single layer media for $R=10^7 \,\mathrm{kOe/s}$ is significantly larger than the corresponding results for $R=7.5 \,\mathrm{kOe/s}$ a comparison of the results presented in in Fig.~\ref{fig:Hs_DeltaE}  show that the normalized values of $H_s(\mathrm{ECC})/H_s(\mathrm{SL})$ track each other reasonably closely for smaller values of ${\Delta E}/k_BT$  but start to diverge at ${\Delta E}/k_BT \approx 42$ $(I\approx1.8\,\mathrm{erg/cm^2})$ where $H_s(7.5 \,\mathrm{kOe/s})$ begins to increase while $H_s(7.5 \,\mathrm{kOe/s})$ drops sharply.

In this work we have presented a formulation of the kMC algorithm that can be successfully applied to simulate dual-layer recording media at the relatively long time scales relevant to experimental MH loops, that are inaccessible using standard LLG micromagnetic simulations.  The results illustrate excellent agreement with a series of experimental measurements on single and dual layer CoPtCrB/CoPtCrSiO granular media. For the first time, a quantitative relation between strength of the interlayer exchange parameter $I$ and spacer-layer thickness $d$ is calculated.  Extraction of the energy barriers relevant for the thermal stability of the recording layer as a function of the saturation field provides a useful figure of merit for ECC media. Such modeling efforts can serve as useful guidance toward the optimization of this important parameter.

This work was supported by Western Digital Corporation, the Natural Science and Engineering Research Council (NSERC) of Canada, the Canada Foundation for Innovation (CFI), and the Atlantic Computational Excellence network (ACEnet).

\bibliographystyle{aipnum}
\bibliography{nanoAPL}

\end{document}